\documentclass[twocolumn,aps,prb,superscriptaddress,showpacs]{revtex4}
\usepackage{amsmath,,amsfonts,latexsym,epsfig}
\usepackage{dcolumn}
\let \Im \relax
\DeclareMathOperator{\Im}{Im}

\begin{document}

\title{Localized magnetic states in Rashba dots}

\author{Mircea Crisan}
\affiliation{Department of Theoretical Physics, University of Cluj,
3400 Cluj, Romania}

\author{David S\'anchez}
\affiliation{Departament de F\'{\i}sica,
Universitat de les Illes Balears, E-07122 Palma de Mallorca, Spain}

\author{Rosa L\'opez}
\affiliation{Departament de F\'{\i}sica,
Universitat de les Illes Balears, E-07122 Palma de Mallorca, Spain}

\author{Lloren\c{c} Serra}
\affiliation{Departament de F\'{\i}sica,
Universitat de les Illes Balears, E-07122 Palma de Mallorca, Spain}
\affiliation{Instituto de F\'{\i}sica Interdisciplinar y Sistemas
Complejos (CSIC-UIB), E-07122 Palma de Mallorca, Spain}

\author{Ioan Grosu}
\affiliation{Department of Theoretical Physics, University of Cluj,
3400 Cluj, Romania}

\date{\today}

\begin{abstract}
We study the formation of local moments in quantum dots arising
in quasi-one dimensional electron wires due to localized
spin-orbit (Rashba) interaction. Using an
Anderson-like model to describe the occurrence of the magnetic
moments in these Rashba dots, we calculate the local magnetization
within the mean-field approximation.
We find that the magnetization becomes a nontrivial function
of the Rashba coupling strength.
We discuss both the equilibrium and nonequilibrium cases.
Interestingly, we obtain a magnetic phase which is stable
at large bias due to the Rashba interaction.
\end{abstract}

\maketitle

\section{Introduction}
   Spin-related phenomena have recently attracted much attention,
    as they are the key ingredient in the new field known
    as spintronics \cite{spintronic}. Two-dimensional (2D) semiconductors are
    appropriate materials to be used in spintronics applications
    since they offer the possibility of an electric control of spins
    via tunable spin-orbit (SO) interaction. An important
    contribution to SO effects in 2D electronic states of narrow
    gap semiconductors (e.g, InAs) is the Rashba interaction \cite{ra}.
    This interaction is a generalization of the vacuum SO
    interaction from the Pauli equation,
    $H_{so}=(e\hbar^{2}/4m^{2}c^{2}){\mathbf{\sigma}}\cdot
    (\mathbf{\nabla} V(r)\times\mathbf{k})$,
    which is small for non-relativistic momenta $\hbar k\ll mc$,
    $V(r)$ being the scalar potential. In semiconductors, the
    energy gap $E_{g}$ and the band splitting $\Delta$ are
    comparable in magnitude ($E_{g}\sim \Delta\sim 1 eV$) and,
    as a consequence, the SO coupling is enhanced by a factor
    $m^{2}c^{2}/E_{g}$.

    The Rashba interaction is a type of SO interaction arising
    when a 2D electron gas forms at the interface of a heterostructure.
    To lowest order in momentum, the Rashba Hamiltonian reads
\begin{equation}\label{eq_ham}
H_{R}=\frac{1}{2\hbar}
([\alpha,p_{y}]_{+} \sigma_{x}-[\alpha,p_{x}]_{+}\sigma_{y})\,,
\end{equation}
   where $\alpha$ is the Rashba coupling proportional to the electric
   field producing the confinement. We take the confinement direction
  along z. In Eq.~(\ref{eq_ham}),
  $\mathbf{p}=(p_{x},p_{y})$ is the 2D momentum and $\sigma_{i}(i=x, y, z)$
   are the Pauli matrices.

   We note that available experimental data \cite{ta} on few-electron
   quantum dots have been discussed in terms of  Rashba spin-orbit
   coupling and exchange interaction \cite{ko}. Using the Spin
  Density Functional Theory it was showed \cite{go} that the
  competition of this coupling and the exchange interaction gives
  rise to the suppression of the Hund rule, and a dot with a closed
  configuration presents a paramagnetic behavior. We have to mention
  that these result have been obtained in the absence of the Coulomb
  interaction.

\begin{figure}
\centerline{
\epsfig{file=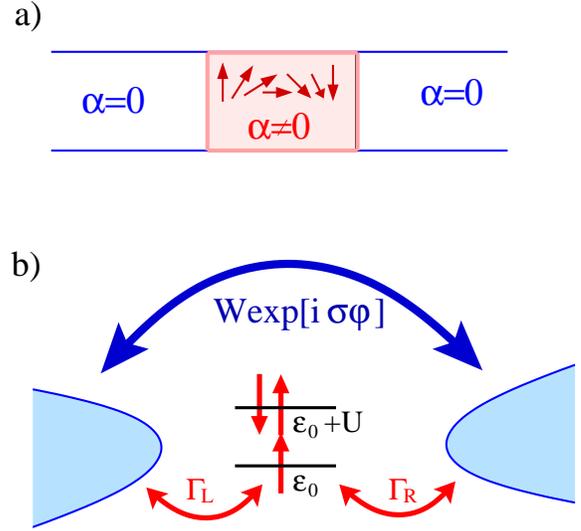,angle=0,width=0.42\textwidth}
}
\caption{(Color online). (a) The system under study consists of a quantum
wire with a region of localized spin-orbit interaction of the Rashba type
($\alpha$ is the Rashba coupling strength). Interactions are restricted to the Rashba dot.
(b) Mapping of the upper system onto a quasi-localized level
 and a nonresonant background channel
with spin dependent couplings to external leads.}
\label{fig_system}
\end{figure}

   When the Rashba interaction is localized around a finite region
   of a quasi one-dimensional ballistic wire [see Fig.\ \ref{fig_system}(a)],
   Ref. \onlinecite{sa1} predicts the formation of quasi-bound states
   which are coupled to the nonresonant background channel.
   Both the potential well and the intersubband coupling
   is produced by the Rashba interaction alone. Furthermore,
   the quasi-bound states lead to enhanced backscattering, causing strong
   dips in the conductance curves of the wire as a function of the
   Fermi energy.\cite{zhang} Since both the level position and broadening can
   be tuned with the Rashba strength $\alpha$, these states are termed
   {\em Rashba dots} \cite{sa1}.

Recently, L\'opez \textit{et al.} \cite{lo} have formulated  a
microscopic theory for transport across Rashba dots
including Coulomb interactions in the dot. An important
aspect of this model is that different regimes can be achieved by tuning
the parameters of the Rashba Hamiltonian and this can be done modulating
external electric fields applied to nearby gates.
The difference between the Anderson Hamiltonian \cite{and}
and the Hamiltonian proposed in Ref. \onlinecite{lo} is twofold.
First, in the Anderson Hamiltonian
the spin is conserved at low temperatures, leading to the Kondo
effect, but in Ref. \onlinecite{lo} the Rashba dot Hamiltonian contains a
spin-flip interaction because the localized states couple
to the continuum states with opposite spins.
Second, due to the Rashba induced precession term,
the direct transmission channel presents
a phase term similar to the Aharonov-Bohm case, but the phase is now
spin-dependent \cite{ac}. Remarkably, despite these differences
the system shows a persisting Kondo effect
at low temperatures but with a novel gate dependence \cite{lo}.

In this paper, we address the magnetic properties of
Rashba dots. We follow Anderson's model for magnetic impurities
in a metallic host and determine whether it is energetically
favorable for the dot to form a localized magnetic moment.
We show below  that the Coulomb
interaction can develop magnetic moments in a Rashba dot for a
critical value of the ratio $(U/\Gamma)_{\rm crit}$, which depends
of the parameters of the Rashba interaction.
Our results might also be important for quantum
dots doped with magnetic impurities \cite{le,fe}. Magnetic ordering in
dots can be induced by the Coulomb interaction and the
magnetization can be electrically controlled even for a fixed
number of electrons \cite{zu1, zu2}. 

 This paper is  organized as follows. We present in section II the
 model and calculate the Green function
using the equation of motion method. The magnetic moment is
 determined in section III both for the equilibrium and the nonequilibrium
cases. The main results are compared
with exact numerical calculations in Sec. IV. The
results are discussed in section V, which also contains
our conclusions. 

   \section{Theoretical Model}

    We start with the model Hamiltonian:
  \begin{equation}
  H= H_e + H_d +H_W+H_V\,,
  \end{equation}
  where
  \begin{eqnarray}
  H_e &=& \sum_{\alpha,k,\sigma } \epsilon_{\alpha,k,\sigma
  }c_{\alpha,k,\sigma}^{\dag}c_{\alpha,k,\sigma}\,, \nonumber \\
  H_d &=& \sum_{\sigma}\epsilon_{d} d_{\sigma}^{\dag}d_{\sigma}
  + Un_{\sigma}n_{-\sigma}\,, \nonumber \\
  H_W&=&\sum_{k,\sigma}(We^{is_{\sigma}\varphi}c_{L,k,\sigma}^{\dag}c_{R,k,\sigma}+H.c.)\,,\nonumber \\
  H_V &=&\sum_{\alpha,k,\sigma}(V c_{\alpha,k,\sigma}^{\dag}d_{-\sigma}+H.c.)\,.
  \end{eqnarray}

In this Hamiltonian we consider the spin quantization axis along
the Rashba field (the $y$ direction for transport along $x$),
$n_{\sigma}=d_{\sigma}^{\dag}d_{\sigma}$ is the
occupation number for electrons in the Rashba dot
with spin $\sigma=\uparrow,\downarrow$ and
$c_{\alpha,k,\sigma}^{\dag} $ is the creation operator of continuum
electrons with wave vector $k$ and spin $\sigma$ in the lead
$\alpha = L,R$. The nonresonant channel
is described with the term $H_W$ where the propagation
phase acquired by a transmitted electron is spin-dependent
($s_{\sigma}=\pm 1$ if $\sigma
=\uparrow,\downarrow$). Finally,
localized and extended electronic states are coupled
via the interaction $H_V$. A pictorial representation
of $H$ is shown in Fig.\ \ref{fig_system}(b).

The parameters of this Hamiltonian are
$U=U(\alpha,l)$, $V=V(\alpha,l)$ and $W=W(\alpha,l)$,  where
$l$ is the length of the Rashba induced square-well potential
(we assume, for simplicity, that $\alpha(x)$ is constant
if $0<x<l$ and zero otherwise) \cite{sa1}, and
$\varphi=k_{R}l$ with $k_{R}=m\alpha/\hbar^{2}$.
Importantly, these parameters can be
externally controlled with gates
by changing $\alpha$ and $l$.

The form of $H$ is similar to the Hamiltonian  
describing the transport in a device formed by an
Aharonov-Bohm interferometer with a quantum dot in one of its arms \cite{ho}
but they differ in that 
the phases in the interaction term $W$ depend on the
spin direction, and that each hopping process through the dot
is associated with a spin-flip event. In the conventional Anderson
model, spin is conserved, and this leads to Kondo correlations.

In order to study the occurrence of magnetic moments in this model we will
calculate the Green function $G_{d\sigma}(\omega)\equiv
\langle \langle d_{\sigma}|d_{\sigma}^{\dag}\rangle \rangle $, which
obeys the equation,
\begin{equation}
\omega G_{d\sigma}(\omega)=1+\langle \langle [d_{\sigma},H]|d_{\sigma}^{\dag}\rangle \rangle \,.
\end{equation}
In the mean-field approximation, the spin-dependent
energy of the $d$-electrons is
$\epsilon_{d,\sigma}=\epsilon_{d}+U\langle n_{-\sigma}\rangle $.
Using the general equation-of-motion method, we find
\begin{eqnarray}
[G_{d\sigma}(\omega)]^{-1}&=&\omega-\epsilon_{d\sigma} +
\frac{1}{\Sigma}\sum_{\alpha k}\frac{2V^{2}}{\omega-\epsilon_{k}}\nonumber\\
&+&\frac{4W\cos\varphi}{\Sigma}\left(\sum_{k}\frac{V}{\omega-\epsilon_{k}}\right)^2\,,
\end{eqnarray}
where $\Sigma$ is given by the expression
\begin{equation}
\Sigma=1-\left(\sum_{k}\frac{W}{\omega-\epsilon_{k}}\right)^2
\end{equation}
If we perform the summations over $k$ in the relations above,
the Green function becomes
 \begin{equation}\label{eq_green}
 G_{d\sigma}(\omega)=\left[\omega-\epsilon_{d\sigma}+i\Gamma+\Gamma\sqrt{x}\cos\varphi\right]^{-1}\,,
 \end{equation}
 where $\Gamma=\Delta/(1+x)$, $x=\pi^{2}W^{2}\nu^{2}$,
 $\Delta=\pi V^{2}\nu$, $\nu$ being the continuum density of states at
 the Fermi level $E_F$ (we take $\nu$ as a constant function of energy).
We note that the tunneling broadening $\Delta$
 becomes renormalized into $\Delta/(1+x)$ due to the background channel
 when $W\neq 0$. Furthermore, the spin contribution due to the Rashba
 interaction is proportional to $\cos\varphi$ for the both spin
 orientations.

\section{Magnetic moment}

\subsection{Equilibrium case}

The magnetization along the Rashba field direction
is given by the difference between the occupancy
expectation value for spin up and spin down,
\begin{equation}\label{eq_magn}
m=\langle n_{d\uparrow}\rangle - \langle n_{d\downarrow}\rangle\,.
\end{equation}
At zero temperature, the occupation reads,
\begin{equation}\label{eqnd}
\langle n_{d\sigma}\rangle =
\int_{-\infty}^{E_{F}}\rho_{d}(\omega)\,d\omega\,,
\end{equation}
where $\rho_{d\sigma}(\omega)$ is the local
 density of states at the Rashba dot, defined in terms
of the Green function as
$\rho_{d\sigma}(\omega)=-\Im G_{d\sigma}(\omega)/\pi$.

Consider first the simple case $U=0$. Then,
 \begin{equation}\label{eqrhod_uzero}
\rho_{d\sigma}(\omega)=\frac{\Gamma}{\pi(\Gamma^{2}+\xi(\omega)^{2})}\,,
 \end{equation}
where $\xi$ is given by
\begin{equation}
\xi(\omega)=\omega-\epsilon_{d}+\Gamma\sqrt{x}\cos\varphi\,.
 \end{equation}
Since $U=0$, the energy $\xi$ is spin independent.
Inserting Eq.~(\ref{eqrhod_uzero}) in Eq.~(\ref{eqnd}),
we obtain
\begin{equation}
\langle n_{d\sigma}\rangle = \frac{1}{2}-\frac{1}{\pi}\tan^{-1}
\frac{\epsilon_{d}-E_F-\Gamma \sqrt{x}\cos\varphi}{\Gamma}\,.
\end{equation}
Because $\langle n_{d\uparrow}\rangle = \langle n_{d\downarrow}\rangle$
even in the presence of spin-orbit interaction,
we trivially have $m=0$. As expected, equilibrium magnetic states
arise due to the presence of Coulomb interactions only.

 Consider now the interacting case $U\neq 0$. We calculate the
 density of states using the spin dependence introduced by
 $\epsilon_{d\sigma}$ and we get:
 \begin{equation}
 \rho_{d\sigma}(\omega)=\frac{\Gamma}{\pi}\frac{1}{\xi_{-\sigma}(\omega)^{2}+\Gamma^{2}}\,,
 \end{equation}
where
 \begin{equation}
\xi_{-\sigma}(\omega)=\omega-\epsilon_{d}+\Gamma\sqrt{x}\cos\varphi-U\langle n_{-\sigma}\rangle\,.
\end{equation}
The occupation reads,
 \begin{equation}\label{eq_ndsigma}
\langle n_{d \sigma}\rangle =\frac{1}{2}-\frac{1}{\pi}\tan^{-1}\frac{\epsilon_{d}-E_F+U\langle n_{d,-\sigma}\rangle-\Gamma \sqrt{x}\cos\varphi}{\Gamma}\,.
\end{equation}

We analyze the formation of a magnetic state from the condition
 \begin{equation}
\frac{d\langle n_{d \sigma}\rangle }{d\langle n_{d,-\sigma}\rangle }=
-U\rho_{d,\sigma}(E_{F})\,.
\end{equation}
As a consequence, the condition for the magnetic state that
 $d\langle n_{d}^{\sigma}\rangle /d\langle n_{d}^{-\sigma}\rangle<-1$ becomes
 \begin{equation}
 U\rho_{d}^{\sigma}(E_{F})> 1\,.
 \end{equation}
  This relation is similar to the Stoner condition for the
  occurrence of the magnetic state in the itinerant-electron
  systems, and the correlations effects appear only as an energy
  shift. A more accurate discussion, taking the energy dependence
  of $\Delta$ (via the density of states $\nu$) changes the magnetic
  region, which is known for a constant density of states.
  However, for our qualitative discussion we follow the wide-band
  approximation with a constant $\Delta$.

  From Eqs.~(\ref{eq_magn}) and~(\ref{eq_ndsigma}) we find a pair
of self-consistent equations for the magnetization $m$
and the total electron density
$n_d=\langle n_{d\uparrow}\rangle +\langle n_{d\downarrow}\rangle $:

  \begin{eqnarray}
  m&=&\frac{1}{\pi}\sum_{\sigma}s_\sigma\cot^{-1}\frac{\frac{U}{2}(n_{d}-s_\sigma m)-\xi(E_F)}{\Gamma} \,, \\
n_d&=&\frac{1}{\pi}\sum_{\sigma}\cot^{-1}\frac{\frac{U}{2}(n_{d}-s_\sigma
m)-\xi(E_F)}{\Gamma}\,,
\end{eqnarray}
From these two equations we calculate the size of the interaction
 above which a local moment develops. On the critical boundary describing
the transition into the magnetic state, we approximately have $m\approx 0$
and $\langle n_{d\uparrow}\rangle\approx \langle n_{d\downarrow}\rangle\approx n_d$.
Thus, we find
 \begin{equation}\label{eq_crit}
\left(\frac{U}{\Delta}\right)_{\rm crit}=\frac{\pi\
(1+c^{2})}{1+x}\,,
 \end{equation}
 where $c=\cot{(\pi n_{d}/2)}$. This condition provides a number of interesting
predictions. First, for increasing $x$ the function $(U/\Delta)_{\rm crit}$ decreases.
Thus, the formation of magnetic moments is enhanced by the coupling to the
continuum states, which is governed by the intensity of the Rashba
interaction. Despite the fact that spin-orbit interactions are
time-reversal symmetric and do not induce spontaneous magnetizations,
indirectly the Rashba coupling makes it more favorable
to generate magnetic solutions as compared to the case without
Rashba interaction. If $x=0$ (or, equivalently, $W=0$)
we recover the condition for the occurrence of the Anderson moments \cite{and}.
Second, $(U/\Delta)_{\rm crit}$ is a weakly
function of the phase $\varphi$ since in Eq.~(\ref{eq_crit})
the dependence on $\varphi$ is only implicit through the total density
$n_d$. Nevertheless, in the general case the condition given by Eq.~(\ref{eq_crit}) 
is far from being trivial since we recall that $U$, $\Delta$ and $n_d$
are complicated functions of the Rashba strength and the dot size \cite{lo}.

\subsection{Nonequilibrium case}

We now turn to the nonequilibrium case, where a finite dc bias $V$
is applied between the two electrodes. The formation of magnetic moments
within the Anderson Hamiltonian out of equilibrium has been recently
analyzed by Komnik and Gogolin, see Ref. \onlinecite{kom}.
They find that the magnetic phase is stable at arbitrarily large voltages
in the case of asymmetric couplings. Here, we assume symmetric couplings
($\Gamma_L=\Gamma_R$) and show below that even in this case the combination
of Rashba interaction and finite bias leads to magnetic moment formation.

At nonequilibrium, the spin-dependent occupations are given by the Keldysh
(lesser) Green function
$G_\sigma^<(t,t')=i\langle d_\sigma^\dagger(t) d_\sigma(t')\rangle$,
\begin{equation}
\langle n_\sigma \rangle=\int_{-\infty}^{\infty}\frac{d\omega}{2\pi}\, 
{\rm Im}\,G_\sigma^<(\omega)
\end{equation}
where $G_\sigma^<(\omega)$ is the Fourier transformed lesser Green function.
We find
\begin{equation}\label{eq_gless}
G_\sigma^<(\omega)=2i\Gamma\frac{\sqrt{x}\sin(s_{\bar{\sigma}}\varphi)(f_L-f_R)/(1+x)
+(f_L+f_R)/2}{\xi(\omega)^2+\Gamma^2}\,,
\end{equation}
where $f_\alpha=1/[1+e^{\beta(\omega-\mu_\alpha)}]$ is the
Fermi distribution function at the lead $\alpha=L,R$ with
inverse temperature $\beta=1/k_B T$.
Interestingly enough, the occupation depends on a term proportional
to $\sin{(\bar{\sigma}\varphi)}$ which has a different sign for opposite spins.
This term appears only at nonequilibrium ($f_L\neq f_R$). Therefore,
we expect a spin polarization ($m\neq 0$) for a noninteracting Rashba dot ($U=0$)
induced by the interplay effect of external bias and Rashba interaction.\cite{sun}

We take $\mu_L=E_F+eV/2$
and $\mu_R=E_F-eV/2$ for the electrochemical potentials in the left
and right contacts. As a result,
we obtain a closed expression for the magnetization,
\begin{equation}\label{eq_mnoneq}
\begin{split}
m&=-\frac{\sqrt{T_r}}{\pi}\sin\varphi\left[\tan^{-1}\frac{eV-2\xi(E_F)}{2\Gamma} \right.\\
&\quad \left. +\tan^{-1}\frac{eV+2\xi(E_F)}{2\Gamma}\right]
\end{split}
\end{equation}
where $T_r=4x/(1+x)^2$ is the background channel transmission.
We infer that the magnetization is negative for positive $V$, arising
from the orientation of the effective Rashba field, which points
along $-y$ \cite{ser08,brus}.
The magnetization can be reversed 
if $\varphi$ changes sign (equivalently, the Rashba intensity $\alpha$).
Obviously, the periodic dependence in Eq.\ (\ref{eq_mnoneq})
arises from the model but it is reasonable to assume a small $\alpha$.
Therefore, $\varphi$ should not be very large. The periodic dependence on $\varphi$ is obtained
in analogy with the Aharonov-Bohm effect and can be found
in related spin-orbit systems (see, e.g., Ref. \onlinecite{sun}).

For $\varphi=\pi/2$ the minimum magnetization reads,
\begin{equation}\label{eq_mmin}
m_{\rm min}=-2\frac{\sqrt{T_r}}{\pi} \arctan\frac{eV}{2\Gamma}
\end{equation}
which approaches $-1$ in the limits of $T_r\to 1$ and $eV\gg 2\Gamma$.
On the other hand, for gate voltages much larger than the applied bias
the magnetization approaches zero as
\begin{equation}
m =-\frac{\sqrt{T_r}}{\pi} \frac{\Gamma eV}{\varepsilon_d^2+\Gamma^2}
\end{equation}
Finally, we note that the magnetization becomes finite and
independent of the gate voltage in the limit of infinite $V$,
in which case we find the simple expression $m=-\sqrt{T_r}$.

In the interacting case, one must replace $\xi(\omega)$
with $\xi_{-\sigma}(\omega)$ in Eq.\ (\ref{eq_gless}).
Then, the expression become involved and the full phase diagram
can be obtained only numerically. However, the model
is tractable in special cases.
In particular, we focus on level energies around
the particle-hole symmetric point ($\varepsilon_d=-U/2$).
We define the dimensionless parameters
$p=-\varepsilon_{d}/U$, $y=U/\Gamma$,
$z=eV/2\Gamma$ and $R=2(\Gamma/U)\sqrt{x}\cos\varphi$.
Our goal is to characterize the critical line
that separates the nonmagnetic and the magnetic phases.
This is given by a curve $p_c$ versus $y_c$ in the
$p$--$y$ plane for different values of $z$.
Then, for small values of $p_c -1/2$ we find
(see the Appendix),
\begin{equation}\label{eq_yc}
y_{c}\approx
\frac{\pi}{2}(1+z^{2})\left[1+(1-3z^{2})
\left[\frac{\pi}{2}(p_{c}-\frac{1}{2})+\frac{R_o}{1+z^{2}}\right]^{2}\right]\,,
\end{equation}
where $R_0=\sqrt{x}\cos\varphi$.
This result shows that the phase diagram presents a dip for
$z>z^\ast$, where $z^{\ast}=1/\sqrt{3}$ (as in Ref. \onlinecite{kom}) but the
form of the phase diagram is modified by the Rashba parameter $R$.
In fact, the dip position shifts away from the symmetric point
due to the Rashba induced level renormalization.
The most important consequence is that whereas
$m$ vanishes for $\varepsilon_d=-U/2$ at large bias in the case
without spin-orbit interactions \cite{kom},
the magnetization remains finite in the Rashba case.
We have numerically checked this prediction (see below).
\begin{figure}
\centerline{
\epsfig{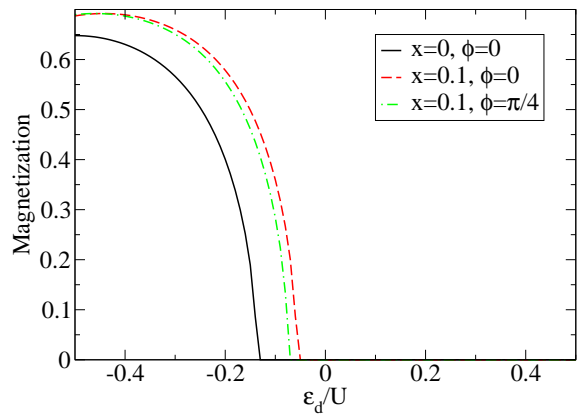}
}
\caption{(Color online). Equilibrium magnetization of the Rashba dot for $\Delta=0.2$
and the values of $x$ and $\varphi$ shown in the figure.}
\label{fig_m}
\end{figure}

\section{Numerical results}

\subsection{Equilibrium case}
We now numerically solve Eqs.~(\ref{eq_magn}) and~(\ref{eq_ndsigma}).
For simplicity, we neglect the dependence of the system parameters on
$\alpha$ and treat $U$, $\Delta$ and $\varepsilon_d$ as independent
constants. In Fig.~\ref{fig_m} we show $m$ as a function of $\varepsilon_d$
for $E_F=0$ and a fixed value of $\Delta$. In the absence of spin-orbit interaction
($x=0$ and $\varphi=0$) the magnetization is zero for positive $\varepsilon_d/U$.
When $\varepsilon_d/U$ decreases, there is a transition point into the
magnetic state, whose magnetization becomes maximal at the particle-hole
symmetric point ($\varepsilon_d=-U/2$). At this point, $m\approx 0.64$,
in agreement with Ref. \onlinecite{and}. We now change the value of $x$ and $\varphi$.
These two parameters can be modified independently tuning $\alpha$
and $l$. Then, for nonzero $x$ and $\varphi=0$ we find that
the transition point shifts toward larger values of $\varepsilon_d/U$.
This results from the self-energy shift $\Gamma \sqrt{x}\cos\varphi$
found in Eq.~(\ref{eq_green}). Moreover, we observe an increase in the amplitude
of $m$ as $x$ increases. This is a consequence of the Rashba coupling enhanced
magnetic moment formation discussed above [Eq.~(\ref{eq_crit})].
Furthermore, keeping $x$ constant and changing $\varphi$ we find
that the magnetization curve changes only slightly, confirming
our earlier prediction.
\begin{figure}
\centerline{
\epsfig{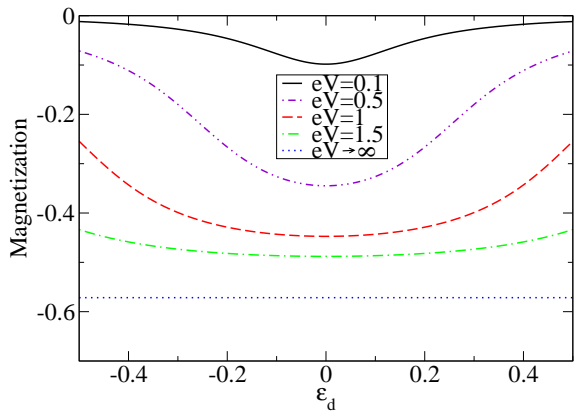}
}
\caption{(Color online). Nonequilibrium magnetization of the Rashba dot for
$\Delta=0.2$, $x=0.1$, $\varphi=\pi/2$ as a function of the level position
for different values of the bias voltage $V$ in the noninteracting case
($U=0$).}
\label{fig_m2}
\end{figure}

\subsection{Nonequilibrium case}
The nonequilibrium magnetization for the noninteracting case ($U=0$)
is shown in Fig.\ \ref{fig_m2} for increasing values of the external
bias $V$. The curves are symmetric around $\varepsilon_d=0$,
which corresponds to the alignment between $\varepsilon_d$ and the Fermi energy.
The magnetization is nonzero for all finite values of $V$,
as discussed after Eq.~(\ref{eq_gless}).
In the limit of large voltages, the curve becomes featureless
according to Eq.~(\ref{eq_mmin}).

\begin{figure}
\centerline{
\epsfig{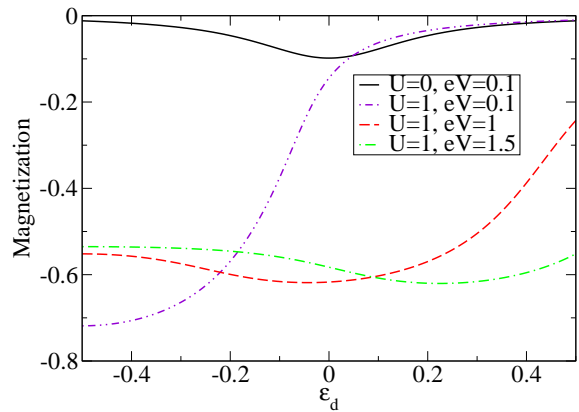}
}
\caption{(Color online). Nonequilibrium magnetization of the Rashba dot for
$\Delta=0.2$, $x=0.1$, $\varphi=\pi/2$ as a function of the level position
for different values of the bias voltage $V$ in the interacting case
($U=1$).}
\label{fig_m3}
\end{figure}

The interacting case is shown in  Fig.\ \ref{fig_m3}.
All energies are given in units of $U=1$.
For comparison, we also reproduce the curve corresponding
to $U=0$ and $eV=0.1$. At the same voltage in the interacting
case, we observe that the magnetization curve follows the noninteracting
curve for large $\varepsilon_d$. This is reasonable since
we are entering the empty orbital regime for which interactions
are unimportant. In the opposite regime, i.e., for negative $\varepsilon_d$,
strong correlations start to dominate
and the interacting magnetization, although still finite,
departs significantly from the noninteracting case. We obtain strong modifications
in the magnetization curve for increasing voltages, favoring the development
of magnetic moments due to the combined influence of interactions
and spin-orbit coupling.
For energies around the particle-hole symmetric point
($\varepsilon_d\approx -U/2$), we find that the magnetization
is reduced as $V$ increases but, unlike the case without spin-orbit interactions,
$m$ does not vanish in the limit of large bias. This is excellent
agreement with the prediction of Eq.~(\ref{eq_yc}).

\section{Conclusions}

We have studied the possibility of the occurrence of the magnetic models in a Rashba dot.
The mean-field approximation has been used to calculate the magnetic moment and the critical
value $U/\Delta$ for the occurrence of the local moments
as function of the parameters of the Rashba Hamiltonian.
This condition, expressed by Eq.~(\ref{eq_crit}),
is similar to the condition obtained for the Anderson model,\cite{and} but contains also
the parameter $x$ which is determined by the Rashba interaction.
Therefore, our calculation suggests a driving of magnetic moments
by external electric fields via the Rashba interaction.
We have demonstrated that the value of the local magnetization
$m$ at equilibrium depends on $x$, but it is worth noting that the curve $m(\epsilon_{d}/U)$
is not very sensitive to the change of the parameter $\varphi$.
This result has been also shown in numerical simulations of the mean-field equations.

As in the standard Anderson calculation,\cite{and}
our mean-field approach breaks the local symmetry
but in an exact solution accounting also for the
effect of the spin fluctuations we should recover the spin rotational invariance.
Nevertheless, even if the magnetic states found above
are an artefact of the model, the mean field solution is interesting as such
since it gives an indication of the
region of the coupling constants of the Hamiltonian where the fluctuations give a relevant effect.
Recently, magnetic moment formation was proposed as a mechanism to explain
the temperature dependence of the conductance
for different gate voltages in quantum point contacts,\cite{me,co} where
the scaling behavior of the conductance close to pinch-off as a function of
temperature was used as an argument for the Kondo effect occurrence.
Hence, our results can be useful for these systems when
spin-orbit interactions become relevant.
We believe that our calculation can also be important for
magnetic semiconductors.\cite{le,ichi}.

In the nonequilibrium regime, we have discussed the interplay between
an external bias and the on-site interaction energy
when the spin-orbit interaction is present.
The phase diagram we obtain is different
from the nonequilibrium case studied in Ref. \onlinecite{kom},
where the spin-orbit coupling was not considered.
In Ref. \onlinecite{kom}, it is shown that
the phase diagram presents a dip for $z>\frac{1}{\sqrt{3}}$.
We have demonstrated that the spin-orbit interaction yields in Eq.~(\ref{eq_yc})
a correction given by the last term proportional
to $R_{0}$. In the case of symmetric model ($2\epsilon_{d}+U=0$) and large bias,
the magnetization vanishes in the absence of spin-orbit interaction.\cite{kom}
In contrast, here we predict that the magnetization remains finite
due to the Rashba interaction. Our numerical solution confirms this result,
which can be particularly relevant for the experiments.
It suggests that in materials with Rashba spin-orbit interaction
the main contribution to the magnetization can be enhanced by applying
a dc bias.

As possible extensions of our model, an interesting possibility is to take into account
an energy dependent density of states (specific for the semiconductors)
like the gapless density of states $\rho(\omega)\sim |\omega|^{r}$.
This will give rise to an energy dependent $\Delta(\omega)$
and the resulting behavior will likely differ from standard quantum dots.\cite{hop}
Future investigations could also deal with the effect of correlations which was neglected in the present calculations.
Using the Hewson decoupling\cite{he} one might follow the method from Ref.\ \onlinecite{bu}
to calculate the effect of magnetic correlations
in the $U\longrightarrow \infty$ limit for systems with spin-orbit interaction.
Finally, progress of experimental studies will be crucial
for the directions development of this model.

\begin{widetext}

\appendix*
\section{}

We present here a derivation of Eq.\ (\ref{eq_yc}).
The mean occupations $n_{\uparrow}$ and $n_{\downarrow}$ have been calculated from
the following relations:
 \begin{align}
2\pi n_{\uparrow}&=-\sqrt{T_r}\sin\varphi\left[\tan^{-1}\frac{eV+2\xi_{\downarrow}(E_{F})}{2\Gamma}+
\tan^{-1}\frac{eV-2\xi_{\downarrow}(E_{F})}
{2\Gamma}\right]+\pi+\tan^{-1}\frac{eV+2\xi_{\downarrow}(E_{F})}{2\Gamma}-\tan^{-1}\frac{eV-2\xi_{\downarrow}(E_{F})}{2\Gamma} \,,\\
2\pi n_{\downarrow}&=\sqrt{T_r}\sin\varphi\left[\tan^{-1}\frac{eV+2\xi_{\uparrow}(E_{F})}{2\Gamma}+
\tan^{-1}\frac{eV-\xi_{\uparrow}(E_{F})}{2\Gamma}
\right]+\pi+\tan^{-1}\frac{eV+2\xi_{\uparrow}(E_{F})}{2\Gamma}-\tan^{-1}\frac{eV-2\xi_{\uparrow}(E_{F})}{2\Gamma}\,.
\end{align}
 Using these results we obtain the magnetization $m$:
\begin{equation}
m=-\frac{1}{2\pi}\sqrt{T_r}\sin\varphi\left[\tan^{-1}\frac{eV+2\xi_{\downarrow}(E_{F})}{2\Gamma}+
\tan^{-1}\frac{eV-2\xi_{\downarrow}(E_{F})}{2\Gamma}+\tan^{-1}\frac{eV+2\xi_{\uparrow}(E_{F})}{2\Gamma}+\tan^{-1}\frac{eV-2\xi_{\uparrow}(E_{F})}{2\Gamma}\right]\nonumber
\end{equation}
\begin{equation}
+\frac{1}{2\pi}\left[\tan^{-1}\frac{eV+2\xi_{\downarrow}(E_{F})}{2\Gamma}
-\tan^{-1}\frac{eV-2\xi_{\downarrow}(E_{F})}{2\Gamma}-
\tan^{-1}\frac{eV+2V\xi_{\uparrow}(E_{F})}{2\Gamma}+\tan^{-1}\frac{eV-2\xi_{\uparrow}(E_{F})}{2\Gamma}\right]
\end{equation}
From this expression we can see that for $U=0$ we have
$\xi_{\uparrow}(E_{F})=\xi_{\downarrow}(E_{F})=\xi(E_{F})$ and the
magnetization $m$ has the value given in Eq. Eq.\ (\ref{eq_mnoneq}) with
$T_{r}=4x/(1+x)^{2}$.

In the same way we calculate the total occupation $n=n_{\uparrow}+n_{\downarrow}$ as
\begin{equation}
n=-\frac{1}{2\pi}{\sqrt{T_r}}\sin\varphi\left[\tan^{-1}\frac{eV+2\xi_{\downarrow}(E_{F})}{2\Gamma}+
\tan^{-1}\frac{eV-2\xi_{\downarrow}(E_{F})}{2\Gamma}-\tan^{-1}\frac{eV+2\xi_{\uparrow}(E_{F})}{2\Gamma}-
\tan^{-1}\frac{eV-2\xi_{\uparrow}(E_{F})}{2\Gamma}\right]
\end{equation}
\begin{equation}
+\frac{1}{2\pi}\left[\tan^{-1}\frac{eV+2\xi_{\downarrow}(E_{F})}{2\Gamma}-\tan^{-1}\frac{eV-2\xi_{\downarrow}(E_{F})}{2\Gamma}+
\tan^{-1}\frac{eV+2\xi_{\uparrow}(F)}{2\Gamma}-\tan^{-1}\frac{eV-2\xi_{\uparrow}(E_{F})}{2\Gamma}+2\pi\right]\,.
\end{equation}
which for $U=0$ becomes
\begin{equation}
n=1+\frac{1}{\pi}\tan^{-1}\frac{8\Gamma\xi}{4\Gamma^{2}-4\xi^{2}+(eV)^{2}}
\end{equation}
for $[(eV)^{2}-4\xi^{2}]/4\Gamma^{2}>-1$. In the limit $V\to 0$ this
equation gives at $E_{F}=0$ and $\varphi=\pi/2$,
\begin{equation}
n=1-\frac{1}{\pi}\tan^{-1}\frac{2\Gamma\varepsilon_{d}}{\Gamma^{2}-\varepsilon_{d}^{2}}\,.
\end{equation}
 From these relations we expect  that  the magnetic ($n_{\sigma}\neq
 n_{-\sigma}$) and non-magnetic solutions
 ($n_{-\sigma}=n_{\sigma}$) exist for small $V$. In the following
 we will analyze the phase diagram taking into consideration the
 extra parameter $V$ and the Rashba interaction. We introduce \cite{and,kom}
the parameter $n_{c}$, which runs from 0 to 1, and
 the dimensionless parameters $p=-\varepsilon_{d}/U, y=U/\Gamma,
 z=eV/2\Gamma$,$R=2(\Gamma/U)\sqrt{x}\cos\varphi$. Thus,
 \begin{equation}
n_{c}=1+\frac{1}{\pi}\tan^{-1}[z+y_{c}(p_{c}+R/2-n+n_{c})]-\frac{1}{\pi}\tan^{-1}[z-y_{c}(p_{c}+R/2-n+n_{c})]\,.
 \end{equation}
Derivating this expression with regard to $n_c$ we arrive at
\begin{equation}
\frac{\pi}{y_{c}}=\frac{1}{1+[z+y_{c}(p_{c}+R/2-n+n_{c})]^{2}}
+\frac{1}{1+[z-y_{c}(p_{c}+R/2-n+n_{c})]^{2}}\,.
\end{equation}
For $p_{c}=1/2$ we can fix $n_{c}=1/2$ and $n=1$,
\begin{equation}\label{eq10}
\frac{\pi}{y_{c}}=\frac{1}{1+(z+R_0)^{2}}
+\frac{1}{1+(z-R_0)^{2}}\,,
\end{equation}
where $R_0=\sqrt{x}\cos\varphi$.
We now write down the equation which contains the small parameter $p_{c}-1/2$,
\begin{equation}\label{eq11}
\frac{\pi}{y_{c}}=\frac{1}{1+[y_{c}(p_{c}-1/2)+z+R_0]^{2}}+\frac{1}{1+[y_{c}(p_{c}-1/2)-z+R_0]^{2}}\,.
\end{equation}
Using Eq.~(\ref{eq10}) we can solve Eq.~(\ref{eq11}) iteratively,
yielding Eq.\ (\ref{eq_yc}). 

\end{widetext}

\section*{Acknowledgments}
This work was supported by the Spanish
Grants No.\ FIS2005-02796 and No. FIS2008-00781,
and the ``Ram\'on y Cajal'' program.

\end{document}